\documentclass{iau} 
\usepackage{graphicx}
\usepackage{subcaption}
\usepackage[square,numbers]{natbib}
\usepackage{amsmath, amssymb}
\usepackage{txfonts}
\usepackage{hyperref}

\title[] %% give here short title %%
{Investigating the impact of different velocity fields on the spectral appearance of Wolf-Rayet stars}

\author[R. R. Lefever et al.]   %% give here short author list %%
{
R.R. Lefever$^1$ $^2$,
T. Shenar$^2$ $^3$,
A.A.C. Sander$^1$, 
L. Poniatowski$^2$, 
K. Dsilva$^1$ \and
H. Todt$^4$
}

\affiliation{$^1$Zentrum für Astronomie der Universität Heidelberg, Astronomisches Rechen-Institut,\\ Mönchhofstr. 12-14, 69120 Heidelberg, Germany\\
  email: {\tt roel.lefever@uni-heidelberg.de}\\[\affilskip]

  $^2$Institute for Astronomy (IvS), KU Leuven, Celestijnenlaan 200D, 3000 Leuven, Belgium\\[\affilskip]

  $^3$Anton Pannekoek Institute for Astronomy, Science Park 904, 1098 XH, Amsterdam, The Netherlands\\[\affilskip]
   
  $^4$Institut für Physik und Astronomie, Universität Potsdam, Karl-Liebknecht-Str. 24/25, D-14476 Potsdam, Germany
}

\pubyear{2022}
\volume{361}  %% insert here IAU Symposium No.
\jname{Massive Stars Near \& Far}
\editors{Jorick Vink}
\begin{document}

\maketitle

\begin{abstract}

\noindent The emission line spectra of WR stars are often formed completely in the optically thick stellar wind. 
Hence, any assumption on the wind velocity law in a spectral analysis has a profound impact on the determination of the stellar parameters.
By comparing Potsdam Wolf-Rayet (PoWR) model spectra calculated with different $\beta$ laws, we show that the velocity field heavily influences the spectra: by using the appropriate $\beta$ laws, the entire range of late and early types can be covered with the same stellar model. 

\keywords{stars: Wolf-Rayet, stars: winds, outflows, stars: atmospheres, stars: mass loss}
\end{abstract}

% \firstsection % if your document starts with a section,
%               % remove some space above using this command.
% \section{Overview}

\begin{figure}[h]
    \centering
    \includegraphics[width=0.7\linewidth]{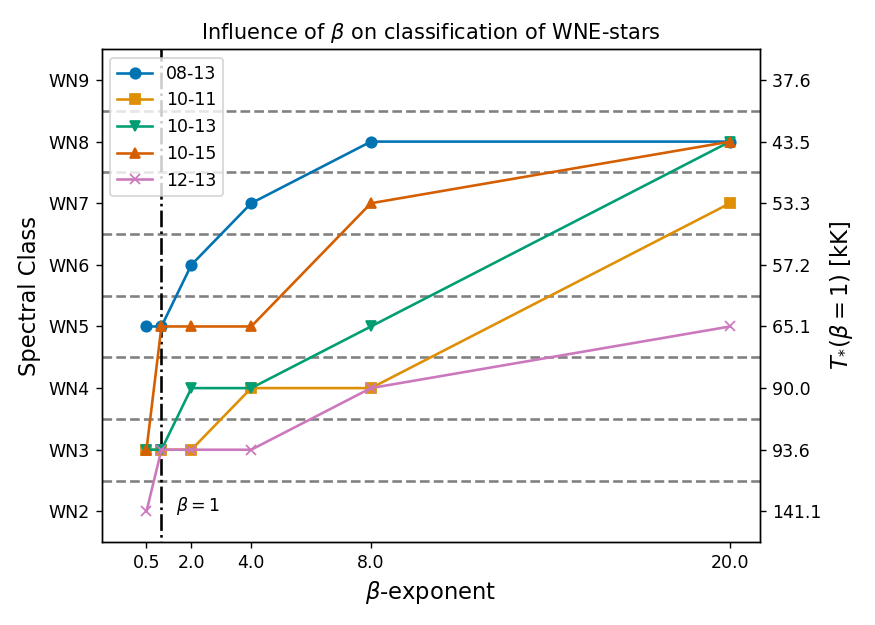}
    \caption{Spectral (sub-)classification for the WN-star models as a function of $\beta$. For labelling we use the same indices as the grids on the \href{https://www.astro.physik.uni-potsdam.de/\~wrh/PoWR/powrgrid2.php}{PoWR website}.}
    \label{fig:classif}
\end{figure}

\noindent Wolf-Rayet (WR) stars play a pivotal role in the energetics of their host galaxies, in the enrichment and ionisation of their local environments and as supernova progenitors. 
Their powerful, line-driven winds form an optically thick regime around the star, causing WR spectra to be dominated by wind features.
For the inference of stellar parameters of WR stars, we rely on model atmosphere codes such as PoWR \citep{grafener2002line, hamann2003temperature, sander2015consistent}. 
To describe the wind velocity field, typically a pre-described velocity law is utilised, the  $\beta$-velocity law \citep[e.g.][]{hamann2006galactic}, which is described as $\varv\,(r) = \varv_\infty \left(1 - \frac{r_0}{r}\right)^{\,\beta}$, with $\varv_\infty$, where for WR stars often $\beta = 1$ is adopted for large samples. \\
\indent When comparing the observed temperatures $T_\mathrm{eff}(\tau_\mathrm{ross} = 20) = T_\ast$ of WR stars from atmosphere modelling to the $T_\mathrm{eff}$ of evolutionary models yielding hydrogen-depleted, core-helium burning stars \citep[e.g.][]{ekstrom2012grids}, there is a clear discrepancy: the observed temperatures are generally considerably lower than the temperatures from evolutionary models due to the extended, optically thick wind. 
Several solutions to solve this problem have been proposed, e.g. via static \citep[e.g.][]{petrovic2006luminous} or dynamical \citep[e.g.][]{grassitelli2018dynamic} inflation. 
Another factor to address here are the $\beta=1$ velocity laws: studies investigating line profile variations \citep[e.g.][]{lepine1999wind} show values of $\beta \sim 5 - 20$, significantly different from a $\beta=1$ assumption for WR stars.\\
\indent To investigate how this $\beta$-parameter influences the spectra, we calculated and compared sets of PoWR model spectra by using different $\beta$ laws.
By investigating diagnostic spectral lines, we see that the $\beta$ parameter has a strong effect on the line spectrum. 
This is illustrated by classifying the spectra of several models using the scheme of \cite{smith1996three}. 
The results for H-depleted WN-star models are shown in Fig. \ref{fig:classif}, with similar results being obtained for WNh and WC stars. 
As a general trend, the model spectra are systematically classified as later-type WR stars for higher $\beta$. 
In fact, by only changing the $\beta$ law, we could cover almost the entire range of spectral types for models with the same $T_\ast$ if we chose $\beta$ values in the range of 0.5 to 20.
As spectral types are typically associated with an apparent temperature, models with an increased $\beta$ parameter would be interpreted as cooler stars.
This is also shown with the average $T_\ast$ per spectral type inferred with $\beta=1$ models on the right axis in Fig. \ref{fig:classif}.
This effect of spectra appearing cooler by using higher-$\beta$ velocity laws causes a degeneracy in the model spectra: by using different $\beta$ laws, models with different $T_{\ast}$, $\dot{M}$ and $\varv_\infty$ can appear similar in the optical spectrum. 
However, this degeneracy can at least in part be solved by P-Cygni lines in the UV as the absorption trough can be used constrain $\varv_\infty$ \citep{prinja1990pcygni}. 

% \section{Conclusions}

% \bibliography{biblio.bib}

\end{document}